\documentclass[12pt,twoside]{article}
\usepackage{amsmath}

\input BoxedEPS
\SetRokickiEPSFSpecial  %% for dvips by Tom Rokicki, for VMS
\HideDisplacementBoxes

\begin{document}
\title{How to mode-lock an atom laser}
\author{Peter D. Drummond$^{1}$, Antonios Eleftheriou$^{2}$,\\ Kerson
Huang$^{2}$,  and Karen V. Kheruntsyan$^{1}$\\
\small $^{1}$Department of Physics, University of Queensland,\\
\small Brisbane, Qld 4072, Australia\\
\small $^{2}$Department of Physics and Center for Theoretical Physics,\\
\small Massachusetts Institute of Technology, Cambridge, MA 02139, USA}
\date{\small MIT-CTP \# 2953}
\maketitle

\pagestyle{myheadings}
\markboth{P.D. Drummond, A. Eleftheriou, K.
Huang,  and K.V. Kheruntsyan}{How to mode-lock an atom laser}
\thispagestyle{empty}

%\pacs{03.75.Fi, 32.80.Pj, 05.30.Jp, 05.45Yv}

\begin{abstract}\noindent
We consider a possible technique for mode-locking an atom laser, based on
the generation of a dark soliton in a ring-shaped Bose-Einstein condensate,
with repulsive atomic interactions. The soliton is a kink, with angular
momentum per particle equal to $\hbar /2$. It emerges naturally, when the
condensate is stirred at the soliton velocity, and subsequently outcoupled
with a periodic Raman pulse-train. The result is a replicating coherent
field inside the atom laser. We give a numerical demonstration of the
generation and stabilization of the soliton.
\end{abstract}
\bigskip

\noindent
The discovery of Bose-Einstein condensation (BEC) in ultra-cold alkali-metal
vapors \cite{BEC} in a magnetic trap, at temperatures of $10^{-6}K$ or
lower, has raised the possibility of a coherent atom laser. Recent
experimental developments \cite{Atom laser} show that this is indeed
practical. All atom lasers to date, however, produce output pulses that are
not phase-coherent. It would be desirable to have a mode-locked laser, in
which the pulses are in phase with each other, for this will enable a wide
range of interference experiments and phase-sensitive measurements. In
addition, mode-locked lasers can have an enhanced intensity stability,
relative to their non-mode-locked cousins.

One way to make a mode-locked atom laser is to create a periodic field
circulating around a ring-shaped condensate, and outcouple it with a
synchronized period. To do this, we must choose a periodic field that can be
easily created, and that has sufficient stability to enable a steady-state
outcoupling process. In this regard, we suggest a dark soliton in a
condensate of atoms with repulsive interactions. This is a kink
configuration, which stands apart from a continuum of possible excitations,
owing to two distinctive features: (a) It has a characteristic propagating
velocity $v_{0}$; (b) The angular momentum per particle, normal to the plane
of the ring, is $\hbar /2.$

The dark soliton can be created by stirring the condensate at the
characteristic velocity $v_{0}$; but it has to be cleaned up, because the
stirring also creates other excitations such as phonons. The cleansing can
be achieve by applying a stroboscopic loss mechanism, which also serves as
outcoupler for the atom laser. The soliton persists because of a dynamical
stability due to the stroboscopic environment, which is particularly strong
for the half-integer angular momentum dark kinks. In the following, we first
describe the soliton as a solution of the nonlinear Schr\"{o}dinger equation
(NLSE) and then demonstrate mode-locking via numerical simulations.

Consider a condensate contained in a ring of radius $R$. Let the
cross-sectional radius be $r_{0}$, and let the cross-sectional area be
denoted by $A=\pi r_{0}^{2}$. We choose $R\gg r_{0}$, so that the transverse
excitations are far more energetic than those along the ring. For the low
excitation modes, therefore, we may regard the condensate as a
one-dimensional system, and denote by $\theta $ the angle around the ring.
The condensate wave function $\Psi (\theta ,t)$ salsifies the NLSE 
\begin{equation}
i\hbar \frac{\partial \Psi }{\partial t}=-\frac{\hbar ^{2}}{2mR^{2}}\frac{%
\partial ^{2}\Psi }{\partial \theta ^{2}}+U\Psi +\frac{4\pi \hbar ^{2}a}{m}%
|\Psi |^{2}\Psi  \label{NLSE}
\end{equation}
where $m$ is the atomic mass, $a$ the $s$-wave scattering length, and $U$ is
an external potential due to trap non-uniformity. The total number of atoms $%
N$ enters through the normalization $AR\int_{0}^{2\pi }d\theta |\Psi |^{2}=N$%
, which is a constant of the motion, if $U$ is independent of the time.
Continuity requires the boundary condition $\Psi (\theta +2\pi ,t)=\Psi
(\theta ,t).$

The NLSE on a line is well-known in nonlinear optics, where it describes the
envelope of an electromagnetic wave propagating along an optical fiber \cite
{optics}. Let us recount what is generally known in the linear case. For
attractive interactions with $\alpha <0$, the nonlinear term in the equation
presents an attractive potential proportional to $|\Psi |^{2}$. In three
spatial dimensions in free space, this would lead to ``self-focusing'' ---
the development of spots of infinite intensity in finite time \cite{collapse}%
. In one dimension, however, the kinetic energy counterbalances the
attraction, leading to the formation of a stable bright soliton. The
situation is the same when we join the ends of the line to form a ring.

For repulsive interactions with $\alpha >0$, we can make a dark soliton in a
linear condensate by requiring $\Psi $ to approach $\pm 1$ at opposite ends
of the line. By continuity, the configuration must have a kink, i.e. a zero
of the wave function. The slope at the\ kink determines its propagation
velocity $v_{0}$, which will be finite and nonzero. Kink solitons in a
cigar-shaped BEC were recently observed \cite{Dark-sol-exp}. Gray solitons
also exist, in which the wave function has a minimum, but never vanishes.
These have also been analyzed theoretically \cite{Pethick}, and created
experimentally \cite{Phillips}. They propagate with a speed proportional to
the wave intensity at the minimum, and hence come to rest in the
dark-soliton limit.

These belong to a different class from the kink soliton we are considering
on a ring, where periodicity demands that the ends match. Therefore, the
phase of the wave function must change by $n\pi $ upon one complete
revolution, where $n$ is an odd integer. This makes the angular momentum per
particle normal to the ring $n\hbar /2$, for the\ same mathematical reason
that an electron has spin 1/2. The lowest-energy dark soliton has $n=1$.

It is convenient to use dimensionless variables to reduce the NLSE to the
form 
\begin{equation}
i\frac{\partial \psi }{\partial \tau }=-\frac{\partial ^{2}\psi }{\partial
\theta ^{2}}+V\psi +\alpha |\psi |^{2}\psi   \label{G-P}
\end{equation}
where $\tau \equiv (\hbar /2mR^{2})t$, $\psi \equiv \Psi \sqrt{{2\pi AR/N}}$,  
$V=(2mR^{2}/\hbar ^{2})U$, and $\alpha \equiv 4NRa/A$. The boundary condition is
$\psi (\theta +2\pi ,\tau )=\psi (\theta ,\tau )$, and the normalization
condition is $\int_{0}^{2\pi }d\theta \psi ^{\ast }\psi =2\pi
$. The energy per particle $\epsilon ,$ and angular momentum per particle $%
\ell$ \ normal to the ring are given by 
\begin{eqnarray}
\epsilon &= &\frac{1}{2\pi }\int_{0}^{2\pi }d\theta \Bigl[ \frac{\partial
\psi ^{\ast }}{\partial \theta }\frac{\partial \psi }{\partial \theta } 
+V\psi ^{\ast }\psi +\frac{\alpha }{2}\left( \psi ^{\ast }\psi \right) ^{2}%
\Bigr]   \nonumber \\
\ell &=&\frac{1}{4\pi i}\int_{0}^{2\pi }d\theta \Bigl( \psi ^{\ast }\frac{%
\partial \psi }{\partial \theta }-\psi \frac{\partial \psi ^{\ast }}{%
\partial \theta }\Bigr)\ .  \label{angmom}
\end{eqnarray}
We consider only special solutions relevant to the application at hand, and
discuss more general solutions in a separate paper \cite{future}.

For configurations that move around the ring at speed $v_{0}$, we denote by $%
\bar{\theta}=\theta -v_{0}\tau $\ the angle in the co-moving frame, and put 
\begin{equation}
\psi (\theta ,\tau )=f(\bar{\theta})\exp \left( i\ell \bar{\theta}-i\omega
\tau \right)   \label{seek soliton}
\end{equation}
where $f$ is real, $\omega $ is a constant to be determined by solving the
equation, and $\ell $ is an integer or half-integer. The boundary condition
requires $f(\bar{\theta}+2\pi )=(-1)^{2\ell }f(\bar{\theta}).$ Substituting (%
\ref{seek soliton}) into (\ref{angmom}), we find that the angular momentum
per particle is $\ell $: 
\begin{equation}
\ell =0,\frac{1}{2},1,\frac{3}{2},\dots.
\end{equation}
Substituting $f$ into (\ref{G-P}), we find the propagation velocity 
\begin{equation}
v_{0}=2\ell\ .
\end{equation}
The equation for\ $f$ is 
\begin{equation}
\frac{d^2f}{d\bar{\theta}^2}=\alpha f^{3}-\beta f 
\label{fequation}
\end{equation}
$\,$where $\beta \equiv \omega +\ell ^{2}.$ Mathematically, this equation
describes the Newtonian motion of a particle of unit mass, in a potential $%
\Phi (f)=-\frac{\alpha }{4}f^{4}+\frac{\beta }{2}f^{2}.$ From the conserved
``energy'' $E=\frac{1}{2}f^{\prime 2}+\Phi (f),$ we can find $f$ in terms of
elliptic functions. Gray solitons correspond to complex $f$, and are not
considered here.

From now on we assume $\alpha >0.$ In the absence of external potential, a
uniform state must have $\psi ^{\ast }\psi =1\,\ $by normalization. These
are the vortex states: 
\begin{eqnarray}
\psi _{\text{vortex}}(\theta ,\tau ) &=&\exp \left[ i\ell \theta -i(\alpha
+\ell )\tau \right]   \nonumber \\
\epsilon _{\text{vortex}} &=&\frac{\alpha }{2}+\ell ^{2}\;\;\;\;\,\text{ \ }%
\left( \ell =0,1,2,3,\dots \right) \ .
\end{eqnarray}
In contrast, the solutions with half-integer $\ell $ must vanish somewhere,
and thus correspond to dark solitons. The width of the soliton should be
proportional to $N^{-1/2},$ for the correlation length of the condensate is $%
(8\pi a\rho )^{-1/2}$, where $\rho $ is the particle density. In the large $%
N $ limit, we obtain the simple form: 
\begin{eqnarray}
f_{\text{soliton}}(\bar{\theta}) &\approx &\tanh \Bigl(\sqrt{\frac{\alpha }{2}}%
(\bar{\theta}-\pi)\Bigr) \nonumber \\
\epsilon _{\text{soliton}} &\approx &\frac{\alpha }{2}+\ell ^{2}+\frac{\sqrt{%
8\alpha }}{3\pi }\;\;\;\;\text{ \ }\Bigl( \ell =\frac{1}{2},\frac{3}{2},%
\frac{5}{2},\dots \Bigr) \,.
\end{eqnarray}
where $\alpha \propto N\gg1$. In this limit we have $\beta \approx \alpha
+(2\pi )^{-1}\sqrt{2\alpha }$. These limiting solitons all have the same
shape, but propagate at different velocities $2\ell .$ Derivation of these
and more general results will be given in a separate paper~\cite{future}.

In experimental situations, the condensate is in equilibrium with a thermal
cloud of uncondensed atoms. The equilibrium number of condensate atoms,
relative to that in the thermal cloud, is determined by the temperature, and
it determines the width of the soliton. During each soliton round-trip, a
loss of particles from the condensate due to output coupling will be
countered by a gain from the thermal cloud. The mean atom number in the
entire BEC is self-stabilizing because the output-coupling efficiency is a
nonlinear function of the dark-soliton width, which in turn is inverse to
the total atom number. Thus, a fluctuation from equilibrium that increases
the atom number, will also increase the efficiency of the output coupler,
which therefore acts as a nonlinear absorber. Even with linear gain and loss
mechanisms, we can create a self-maintained soliton and a steady stream of
coherent output pulses. The feasibility of the scheme can be demonstrated by
numerical simulations, as we now describe.

First, we create a dark soliton with $\ell =1/2$ from a uniform static
condensate, by momentary stirring it at the soliton velocity $v_{0}=1$. In
practice, this can be produced by a blue detuned laser beam. In the
simulation, we introduce a repulsive potential $V(\theta ,\tau )$ to create
a moving ``hole'' in the condensate at $\theta =\tau -\pi $. The time origin
is displaced by $\pi $, to wait for the hole to fully form. Next, we have to
clean up the configuration by adding gain and loss mechanisms, for the
stirring creates other excitations in addition to the dark soliton. The
entire procedure is contained in the generalized equation of motion 
\begin{equation}
i\frac{\partial \psi }{\partial \tau }=-\frac{\partial ^{2}\psi }{\partial
\theta ^{2}}+\alpha |\psi |^{2}\psi +V(\theta ,\tau )\psi +i\left[ g-\gamma
(\theta ,\tau )\right] \psi  \label{G-P+gain-loss}
\end{equation}
where the stirring potential is chosen as: 
\begin{equation}
V(\theta ,\tau )=V_{0}\exp \Bigl[ -\Bigl( \frac{\theta -\tau +\pi }{\sigma
_{1}}\Bigr) ^{2}-\Bigl( \frac{\tau -\pi }{\sigma _{2}}\Bigr) ^{4}\Bigr]\ .
\end{equation}
In the gain-loss mechanism, $g$ is a constant gain rate, representing
stimulated emission from non-condensed atoms that are continuously loaded
into the trap. The loss function makes localized periodic hits at the center
of the soliton: 
\begin{figure}[bt]
\centerline{\BoxedEPSF{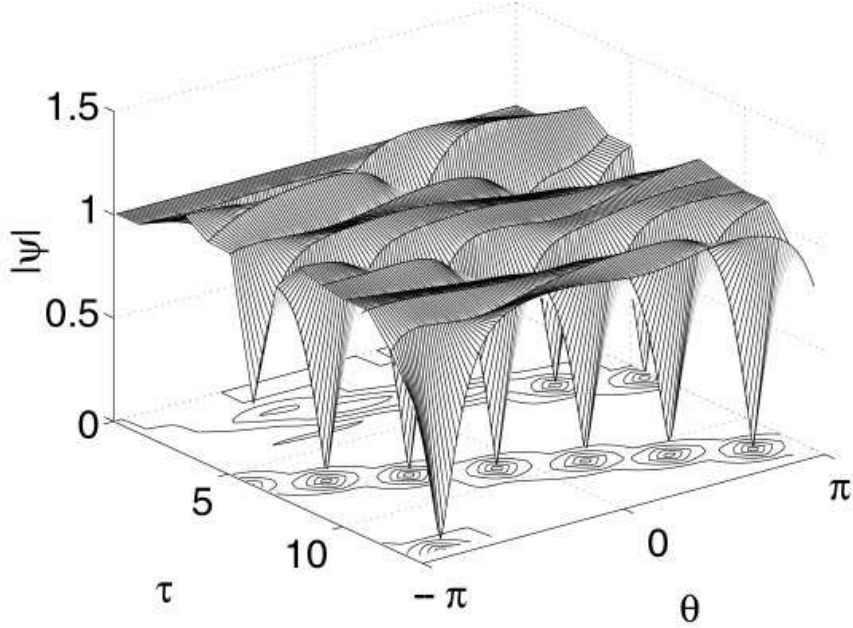 scaled 600} }
\caption{Modulus of the condensate wave function as function of angle around
the ring, and dimensionless time. Stirring an initially uniform condensate
at the soliton velocity creates a moving hole, which develops into a dark
soliton, after unwanted excitations are cleansed by the loss mechanism.}
\label{KH:fig:1}
\end{figure}
\begin{figure}[bt]
\centerline{\BoxedEPSF{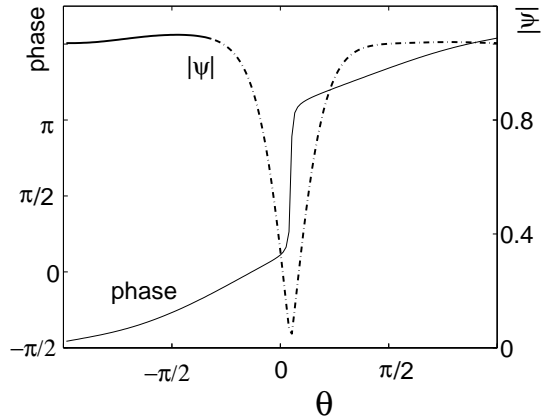 scaled 366}}
\caption{Modulus and phase of the dark soliton, as functions of the angle
around the ring, at a time when it has become fully stabilized. The phase
jumps by $\pi $ across the dip, indicating that the soliton is a kink.}
\label{KH:fig:2}
\end{figure}
\begin{equation}
\gamma (\theta ,\tau )=\gamma _{0}\exp \Bigl[ -\Bigl( \frac{\theta }{\sigma
_{\theta }}\Bigr) ^{2}-\Bigl( \frac{\tau -\tau _{1}}{\sigma _{\tau }}%
\Bigr) ^{2}\Bigr]   \label{loss}
\end{equation}
where $\tau _{1}=$ $\left( \tau \text{ mod }2\pi \right) -\pi $ . This
simulates a stroboscopic output coupler, realizable, for example, through a
local Raman-tuned transition to a non-trapped state of the atoms. The strobe
acts for the first time at $\tau =2\pi $, allowing time for the hole to form.

\begin{figure}[bt]
\centerline{\BoxedEPSF{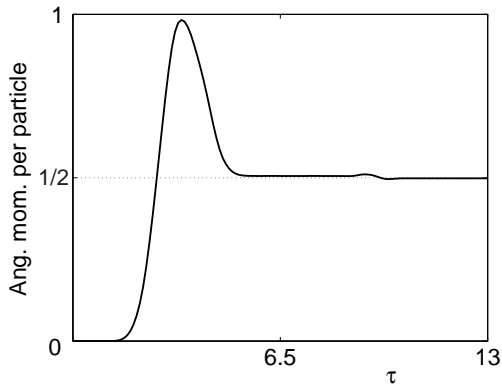 scaled 350}}
\caption{The angular momentum went through transients while the condensate
was being stirred, but stabilizes to a value $\hbar $/2 characteristic of a
kink, under actions of the gain-loss mechanism.}
\label{KH:fig:3}
\end{figure}

In the numerical analysis, we set $\alpha =7.3634$, chosen to correspond to $%
\beta =8$ in (\ref{fequation}). For the stirring potential we use $%
V_{0}=1.673\alpha $ , $\sigma _{1}=1.575\sqrt{2/\beta }$, and $\sigma
_{2}=\pi /2$. The gain and the loss parameters are set to: $g=0.01$, $\gamma
_{0}=0.9$, $\sigma _{\theta }=1.05\sqrt{2/\beta }$, $\sigma _{\tau }=1.05%
\sqrt{2/\beta }$.

\begin{figure}[bt]
\centerline{\BoxedEPSF{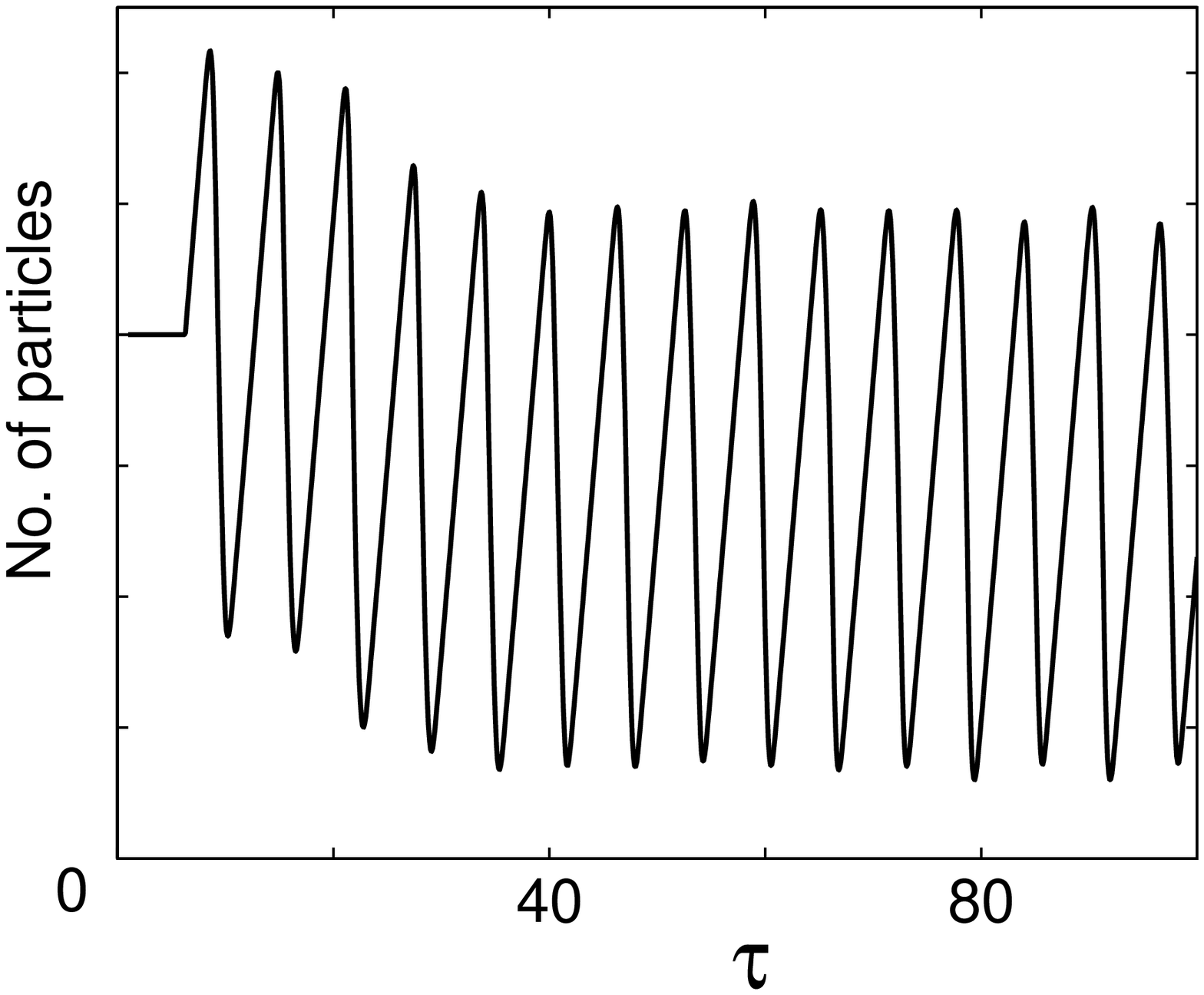 scaled 400}}
\caption{Mode-locking --- production of a coherent train of pulses --- is
indicated by the steady-state oscillation of the total number of atoms in
the condensate.}
\label{KH:fig:4}
\end{figure}

Following are the results of numerical calculations, with the initial
condition $\psi =1$. Fig.~\ref{KH:fig:1} shows $|\psi |$ as a function of $\theta $
and $%
\tau $. We see that a hole was formed, and developed into a dark soliton
through the action of the gain-loss mechanisms. A more detailed view is
shown in Fig.~\ref{KH:fig:2}, which shows $|\psi |$ and the phase of a clean
soliton, as a function of $\theta $ at a fixed time. The signature is that the phase
jumps by $\pi $ across the soliton. The modulus does not precisely vanish
due to small admixtures of non-soliton excitations. Fig.~\ref{KH:fig:3} shows the
angular momentum per particle, which rises from 0 to almost 1 as the condensate is
being stirred, but settles down to $\hbar /2$ characteristic of a kink$,$
under actions of the gain-loss mechanism. Fig.~\ref{KH:fig:4} shows the total number
of atoms in the ring as a function of time. The steady-state oscillations
indicate an output train of pulses coherent with each other. In this
simulation, the stroboscope performs the dual task of \ cleaning up the
soliton, and acting as outcoupler for the mode-locked laser. In general,
these tasks could be done by separate mechanisms.

Our numerical results are quite sensitive to the exact strength of the
inter-atomic repulsive potential and to the gain and loss parameters. In
actual experiments, however, one can adjust the potential, either by taking
advantage of the similar couplings that exist in two-component Bose gases 
\cite{Inter-species}, or by using tuning techniques involving Feshbach
resonances\cite{Feshbach}, which have been recently demonstrated
experimentally\cite{Feshbach-experiment}. It may also be feasible to utilize
inelastic atom-atom collisions as a saturation mechanism, to further
increase the mode-locked laser stability.

We would like to acknowledge useful discussions with M. Holland. Two of the
authors (A.E. and K.H.) were supported in part by a DOE cooperative
agreement DE-FC02-94ER40818. P.D. and K.K. acknowledge ARC for the support
of this work.

%\bigskip

\end{document}